\documentclass[english,aps,pra,reprint,noshowpacs,superscriptaddress]{revtex4-1}   
\usepackage[T1]{fontenc}	% should generally be included for better accented-word behavior
\usepackage[latin9]{inputenc}	% should generally be included for better accent behavior
\usepackage{geometry}		% for controlling page margins
\usepackage{graphicx}
\usepackage[above,below]{placeins}	% allows use of \FloatBarrier command to force section barriers
\usepackage{times}
\usepackage[colorinlistoftodos]{todonotes}
\usepackage{hyperref}  
\usepackage{enumerate}
\usepackage{enumitem}
\usepackage{enumitem}

\hypersetup{colorlinks=true,urlcolor=blue,citecolor=blue,linkcolor=blue}   
\begin{document}

\title{Using reflections to explore student learning during the project component of an advanced laboratory course}
\author{Bei Cai}
\affiliation{Department of Physics, Engineering Physics and Astronomy, Queen's University, 64 Bader Lane, Kingston, ON K7L 3N6, Canada}
\author{Lindsay Mainhood}
\affiliation{Faculty of Education, Queen's University, 511 Union St W, Kingston, ON K7M 5R7, Canada}
\author{Robert G. Knobel}
\affiliation{Department of Physics, Engineering Physics and Astronomy, Queen's University, 64 Bader Lane, Kingston, ON K7L 3N6, Canada}

\begin{abstract}

We redesigned an advanced physics laboratory course to include a project component. The intention was to address learning outcomes such as modeling, design of experiments, teamwork, and developing technical skills in using apparatus and analyzing data. The course included experimental labs in preparation for a six-week team project in which students designed and implemented a research experiment. The final assignment given to students was a reflective essay, which asked students to discuss their learning and satisfaction in doing the project. Qualitative analysis of the students' reflections showed that the majority of the students reported satisfaction and achievement, functional team dynamics, learning outcomes unique to this experience, practicing modeling skills, and potential future improvements. We suggest that reflections are useful as support for student learning as well as in guiding curricular improvements. Our findings may be useful for other course redesign initiatives incorporating project-based learning and student reflections.  
\end{abstract}

\maketitle
\section{Introduction}
\vspace*{-0.15in}
The laboratory has long been an essential part of the undergraduate physics curriculum.  At Queen's University and throughout the physics education community, the redesign and study of undergraduate laboratory courses has become an area of increasing focus~\cite{cite:Schumacher, cite:ZwicklGoals, cite:Dounas1z}. 
The American Association of Physics Teachers (AAPT) recently advised that the learning outcomes of students in physics undergraduate laboratory courses should include the ability to pose research questions, model systems, design experiments, and analyze data~\cite{cite:AAPT}. These outcomes, however, are difficult to achieve when students do procedural experiments in which they follow a well-planned recipe to obtain predictable results. One effective way to achieve the recommended learning outcomes, and to encourage student engagement, is to incorporate student-led open-ended experimental projects.

Our shift toward design- and project-based laboratory courses is motivated not only on the basis of the above recommendations, but also because there is evidence that this mode of instruction changes students' attitudes about physics. For example, lab courses with research and design as the focus have been shown to be effective at increasing student retention in science, technology, engineering, and mathematics (STEM)~\cite{cite:STEMretention}. Lab courses that carefully consider learning goals and the methods of instruction may significantly influence students' scientific practices and attitudes~\cite{cite:Etkina2, cite:Wilcox}. Further, modeling can be used to ``integrate sophisticated conceptual and quantitative reasoning into the experimental process ... a natural way to integrate an analysis and discussion of systematic error into a lab activity''~\cite{cite:ZwicklGoals}. Dounas-Frazer et al. recently found ``concrete implications for the design of experimental physics projects in courses for which student ownership is a desired learning outcome''~\cite{cite:Dounas1z}.

Reflective writing exercises are a key step in metacognition, allowing the student to interrogate their own learning and laboratory process. Student reflections are a benefit for both assessing and supporting students' learning in laboratories~\cite{cite:NAS, cite:Gandhi2016}. Such reflection exercises have been shown to help students develop problem solving skills~\cite{cite:reflection1}, content knowledge~\cite{cite:reflection2}, conceptual understanding~\cite{cite:reflection3, cite:dounas-frazer}, and attitudes about physics~\cite{cite:reflection4}.

In this paper we begin by describing the lab course and the student reflections upon which this research is based. Our redesign efforts in incorporating design and project elements into this lab course are briefly described. We then discuss in depth the findings of our qualitative analysis of the student reflections at the end of their project. Our findings offer some additional implication to existing literature of benefits of design- and project-based lab courses.

\vspace*{-0.15in}
\section{Course Context}
\vspace*{-0.15in}
Our third-year engineering physics laboratory course introduces concepts of quantum and modern physics (nuclear, particle, atomic and optical physics) with increasingly sophisticated apparatus and analysis.  The class of $\sim$60~students meets for a 50-minute lecture and a three-hour lab (in two sections) weekly for 12~weeks.  For several years the course offered four common experiments done in pairs for the first few weeks, followed by a rotation through a selection of experiments, each completed in the three-hour period.  Students reported dissatisfaction with the amount of work required for these procedural experiments, did not demonstrate much retention of the learning goals in subsequent courses, and seemed less engaged than in parallel engineering design project courses. 

The course was substantially redesigned for Fall 2017 with the goal of addressing more of the learning outcomes from the AAPT report~\cite{cite:AAPT}.  To this end, students were given six~weeks to complete an experimental project in teams of four.  The project required students to research an existing experiment of their choosing, design the apparatus and procedure, predict the results using a model, carry out the experiment and analysis, and report on their findings.  In order to help ensure success in such a short time, students were required to find an experiment already published in an undergraduate-level journal, and instructors carefully vetted projects for both feasibility and ambition at the proposal stage.  Projects that the students carried out in Fall 2017 included:  measuring Faraday rotation in water, measuring the speed of light, measuring the angular dependence of cosmic ray muon flux,  determining the abundance of potassium-40 in everyday objects, and measuring the Hall coefficient of aluminum.

The first six~weeks of the course included exercises and experiments aimed at teaching students technical skills and scientific attitudes that are necessary for, and ease the transition to, the open-ended project phase.  There were activities which demonstrated flexible equipment available for later re-use, experiments that focused on data analysis, and exercises that supported core elements of experimental design. %These experiments in the first six weeks, together with the experimental projects, addressed a wide range of the learning outcomes as suggested by the AAPT report.  
In order to evaluate the first delivery of this redesigned course, we used a mixed-methods approach. We took quantitative data with the E-CLASS survey~\cite{cite:ECLASS} to evaluate student attitude and with the LOPUS lab observation protocol~\cite{cite:LOPUS} to evaluate student engagement in the labs. The findings of the quantitative analysis will be reported in a future publication.

As the last deliverable for the course, after presenting their final project report orally and in written form, students were asked to write a short 300-500 word guided reflection essay.  In this essay (worth 3\% of the course grade) students were asked to reflect on the learning and process during the project.  The students were given a rubric showing that this grade came from clarity of writing, reflection on learning, reflection on team, and plans for the future.  Question prompts were given to the students asking them to reflect on: their happiness with what their team achieved; how their team worked during the project; what they learned over the course of the project; what they learned specifically when they compared a prediction to a model; and finally what they would do differently if they were to do the same project again.  One possible confounding effect is that students may report more positive reflections in the hope of obtaining higher grades.

\vspace*{-0.15in}
\section{Methodology}
\vspace*{-0.15in}
Our qualitative approach to explore the learning experiences of students used the text-coding steps (open, axial, and selective coding) described by Corbin and Strauss~\cite{cite:coding}. These analysis steps allowed us to deduce how students' reflection responses addressed our research questions. Importantly, the following five research questions directly informed the reflection prompts given to students, and also guided our analysis of the reflections:
\begin{enumerate}[label=\Alph*., noitemsep,nolistsep]
\item Were students satisfied with the outcome of their project?
\item What team dynamics did the students encounter in their project?
\item What learning did the students describe in their reflection on their project experience?
\item What did the students learn specifically from comparison of model to data?
\item What would the students do to improve their experience next time?
\end{enumerate}

The primary coder (BC) worked closely with the course instructor (RGK) in the course redesign before and during the time the course was offered. BC completed the open coding phase of the student reflections in consultation with LM. Together, BC and LM completed the axial and selective coding phases to create categories and themes.
 
\vspace*{-0.15in}
\section{Results}
\vspace*{-0.15in}
In this section we describe the findings of our qualitative  thematic analysis of the student reflections. Specifically, we discuss the five themes that emerged from the deductive analysis and corresponded to the prompts we provided the students prior to completing the reflection assignment. We discuss the themes separately in five subsections: students' project satisfaction, team dynamics, described learning, experiences with modeling, and future improvements to the project experience. 

\vspace*{-0.15in}
\subsection{Project satisfaction}
\vspace*{-0.15in}
The students were asked whether they were happy with what their team achieved at the end of their experimental project. Out of the 55~responses we received and analyzed, 30~students explicitly wrote that they found their project interesting, regarded their project as a success, and/or were happy with what their team achieved. Three students explicitly reported that they were unhappy, disappointed, or stressed. Six~students wrote that they had mixed feelings: on one hand they were happy with their experimental process but on the other hand they were unsatisfied with their project outcome. The remaining 16~students did not report their level of project satisfaction.

Although we did not prompt the students to do so, 27~students reflected on how they benefited from this course. Their responses conveyed an important message: students valued the project experience because of the unique learning that happened through doing experimental projects. Some example quotations include:

\vspace*{-0.05in}
\begin{quote}
\textit{Being able to delve into this project from start to finish, as opposed to simply taking what is given and testing what is expected, such as in the past, has been a great experience and one which will help me in future endeavours to succeed in collaboration, in sound process, and in design.} 

\textit{I think the design portion of the project is very important as it is the first time I have had to solve physics problems without a known solution, and use the result to produce a meaningful outcome. This put all the physics I've learned over my university career in perspective, and now see how it could all actually be applied.}
\end{quote}
\vspace*{-0.05in}

\vspace*{-0.15in}
\subsection{Team dynamics}
\vspace*{-0.15in}
Most project teams felt that they functioned quite well. Common functional team dynamics included: having clear team expectations, good communication among team members, fair task distribution, friendly relationships that are supportive and trustworthy, and all members being willing to take responsibility and contribute more or less equally. A designated leader was a feature of some functional teams, while other teams felt effective without a defined leader. Students who felt they were part of a functional team said:

\vspace*{-0.05in}
\begin{quote}
\textit{We agreed at the start of our project what our goals and expectations were in terms of time and effort commitments. This ensured that everybody was prepared for the process of our project up to its completion.}

\textit{When there was a task at hand, more than one person volunteered to do it, and it was assigned to the most appropriate individual based on related works or past experience. ... Proper team communication was very important as many in depth topics read by other members were summarized to convey the key points relevant to the lab.}
\end{quote}
\vspace*{-0.05in}

A small fraction of the teams did not feel that they worked well together. Common non-functional team dynamics were described as poor communication, unfair task distribution, and/or lack of responsibility among team members.

\vspace*{-0.05in}
\begin{quote}
\textit{There was a gap in understanding between team members as each member only had a clear understanding of their portion of the project.}

\textit{While no particular group member was specifically bad, the individual sense of responsibility when it came to doing the less exciting part of the projects (background research, writing reports, doing presentations) was lacking.}
\end{quote}
\vspace*{-0.05in}

It is interesting to note that some teams felt functional because they had previous experience working with their team members, this allowed them to \textit{``work off each other's strengths to the best of [their] abilities''}. Yet another student wrote that their non-functional team \textit{``dynamic was caused by having a group of close friends working together, which lowered the pressure of showing up very late to meetings or not doing the required work by the time it is due''}. While the students may feel more comfortable working on their project in the teams that they form themselves, it might not necessarily lead to functional team dynamics. 

\vspace*{-0.15in}
\subsection{Described learning}
\vspace*{-0.15in}
Students were asked to reflect on the prompt, ``What did you learn over the course of the project?'' They reported three major learning items in their reflections: \textbf{ways of thinking, learning by doing projects, and requirements for success}.

The ways of thinking that students felt they learned or developed include conceptual understanding, problem solving, and critical thinking. While students had to have a basic understanding of physics concepts behind their project goal before they started designing their experiment, they felt they practiced conceptual thinking as they progressed through their project phases. In terms of problem solving, students described having to \textit{``figure out the problems presented to [them] and rise above them to complete [their] project''}. They also reported that doing a project allowed them to \textit{``critically think[ing] about results''}. The conceptual understanding, problem solving, and critical thinking skills that students reported learning through the project affirm what value design- and project-based learning experiences afford physics and engineering students. As one student wrote:

\vspace*{-0.05in}
\begin{quote}
\textit{Since it was not the same as following the procedures on an experiment handbook, we had to solve all the problems ourselves, and it really made me think about how each part of the apparatus worked, and it turned out to be essential in the data and error analysis.}
\end{quote}
\vspace*{-0.05in}

The other types of learning reported by students were related to learning by doing projects and their requirements for success. The students wrote that they learned the whole research process including reading literature, modeling, designing experiments, and interpreting data and results. Technical skills and an acquired scientific attitude are byproducts the students felt they gained during their project experience. Students also reflected on what elements were imperative to their project's success. Requirements for success included research skills, project management, working in effective teams, utilizing resources, time management, and making necessary adjustments when difficulties and problems occur. An example of one student's awareness of such needs was described as:

\vspace*{-0.05in}
\begin{quote}
\textit{Through our successes (and failures) as a team, we learned that performing an experiment requires a lot of careful documentation, thoughtfulness regarding procedure and analysis, and excellent communication between team members.}
\end{quote}
\vspace*{-0.05in}

\vspace*{-0.15in}
\subsection{Experiences with modeling}
\vspace*{-0.15in}
The course instructor emphasized from the first description of the project the importance of modeling to the students and explicitly stated that one goal of the project was to compare a prediction with a model.  Despite such efforts to emphasize modeling, less than half of the students reflected on their experiences with modeling. These students reported that they learned how to model but offered no further reflections, and they were able to identify some limitations of their models. There are two possible explanations: the lack of student responses related to modeling may be because that the students did not understand what the instructor wanted them to reflect upon; the lack of depth in reflection on modeling signifies that, as instructors, we can do a much better job supporting students thinking about modeling and developing their modeling skills such as by using the Modeling Framework for Experimental Physics~\cite{cite:ZwicklModel, cite:DounasModel}. 

\vspace*{-0.15in}
\subsection{Future improvements}
\vspace*{-0.15in}
Students reflected on the numerous issues they encountered prior to reflecting on what future improvements could be made. The challenges and issues the students encountered included: coming up with a research question that is appropriate for the short timeline, poor physics conceptual understanding, technical challenges including apparatus and lack of other resources, ineffective team dynamics, poor time management, and obtaining poor experimental results. Without this deeper understanding of students' challenges, which we gained through their reflections, improving the project experience for students in the future would lack consideration of students' learning needs.

Three ideas emerged from student reflections about what future improvements could help them to have a more successful project experience: \textbf{better project management, utilizing resources, and better research methods}. Students voiced that project management is key; the team should consider the whole project phase, properly plan the procedures to allow time for critical tasks, ensure good documentation of steps taken, and work together as a team toward its project expectations. Students recognized that there were resources provided that were not necessarily utilized. They reported that finding out what resources are available and acquiring the appropriate hardware/software would improve the experience. Students also realized that lab courses are normally supported by a team of experts including the course instructor(s), teaching assistants, and technicians, all of whom would be helpful to access. Last but not least, as many of the students wrote in their reflections, doing projects is similar to going through an authentic research process. They reported the need to empower themselves with better research methods, specifically mentioning improved literature search skills, conceptual understanding, scientific attitude, technical knowledge, and data analysis skills. 

\vspace*{-0.15in}
\section{Conclusion}
\vspace*{-0.15in}
In this paper we sought to illustrate the use of and analytic findings from a reflection assignment given to third-year engineering physics students after completing a six-week experimental project. We conclude by suggesting that design- and project-based learning experiences are powerful tools to help students achieve the learning outcomes recommended by the AAPT~\cite{cite:AAPT}. This suggestion is based on our findings that most students were satisfied with their experimental project and achievements, experienced functional team dynamics, described numerous learning outcomes unique to this experience, learned some modeling skills, and were able to assess how an improved experience could better support their learning. These student impressions are consistent with those of the instructional team.  The instructors were uniformly impressed with the increased engagement from the students during the project, by the ambition they showed in the designing of the experiments, and by the satisfaction students took in presenting their final results.  Further planned improvements to the course will attempt to lessen the stress some students reported by emphasizing experimental process over data. We intend for our findings to be useful for other course redesign initiatives incorporating project-based learning and student reflections. 

%\vspace*{0.076in}
\vspace*{-0.12in}
\acknowledgments{This work is funded by the TRESTLE network~\cite{cite:TRESTLE} (NSF DUE1525775), and the Department of Physics, Engineering Physics and Astronomy, the Center for Teaching and Learning, and the Faculty of Engineering and Applied Science at Queen's University.}

\end{document}